\documentclass{article}
\usepackage{amsmath,graphicx}
\usepackage{cite}
\usepackage{algorithmic}
\usepackage[ruled,vlined,commentsnumbered]{algorithm2e}
\usepackage{balance}
\usepackage{subfigure}
\usepackage[preprint]{spconf}

\copyrightnotice{\begin{tabular}[t]{@{}l@{}}© 2020 IEEE.  Personal use of this material is permitted. Permission from IEEE must be obtained for all other uses, in any current or\\future media, including reprinting/republishing this material for advertising or promotional purposes, creating new collective works, for\\resale or redistribution to servers or lists, or reuse of any copyrighted component of this work in other works.\end{tabular}}

\title{GLAUCOMA DETECTION FROM RAW CIRCUMPAPILLARY OCT IMAGES USING FULLY CONVOLUTIONAL NEURAL NETWORKS}
\name{Gabriel Garc\'{i}a$^{1}$, Roc\'{i}o del Amor$^{1}$, Adri\'{a}n Colomer$^{1}$, Valery Naranjo$^{1}$ \thanks{This work has been funded by GALAHAD project [H2020-ICT-2016-2017, 732613], SICAP project (DPI2016-77869-C2-1-R) and GVA through project PROMETEO/2019/109. The work of Gabriel Garc\'{i}a has been supported by the State Research Spanish Agency PTA2017-14610-I. We thank NVIDIA Corporation for the donation of the Titan V GPU used here.}}
\address{$^{1}$Instituto de Investigaci\'on e Innovaci\'on en Bioingenier\'ia (I3B),\\ Universitat Polit\`ecnica de Val\`encia, Camino de Vera s/n, 46022, Valencia, Spain.}

\begin{document}
%
\maketitle
\begin{abstract}
Nowadays, glaucoma is the leading cause of blindness worldwide. We propose in this paper two different deep-learning-based approaches to address glaucoma detection just from raw circumpapillary OCT images. The first one is based on the development of convolutional neural networks (CNNs) trained from scratch. The second one lies in fine-tuning some of the most common state-of-the-art CNNs architectures. The experiments were performed on a private database composed of 93 glaucomatous and 156 normal B-scans around the optic nerve head of the retina, which were diagnosed by expert ophthalmologists. The validation results evidence that fine-tuned CNNs outperform the networks trained from scratch when small databases are addressed. Additionally, the VGG family of networks reports the most promising results, with an area under the ROC curve of 0.96 and an accuracy of 0.92, during the prediction of the independent test set.
\end{abstract}

\section{Introduction} \label{sec: Introduction}

Glaucoma has become the leading cause of blindness worldwide, according to \cite{jonas2018}. It is characterized by causing progressive structural and functional damage to the retinal optic nerve head (ONH). Recent studies advocate that roughly 50\% of people suffering from glaucoma in the world are undiagnosed and ageing populations suggest that the impact of glaucoma will continue to rise, affecting 111.8 million people in 2040 \cite{reference2_intro}. Therefore, early treatment of this chronic disease could be essential to prevent irreversible vision loss. 

\begin{figure}[b]
\centering
\includegraphics[width=8.5cm]{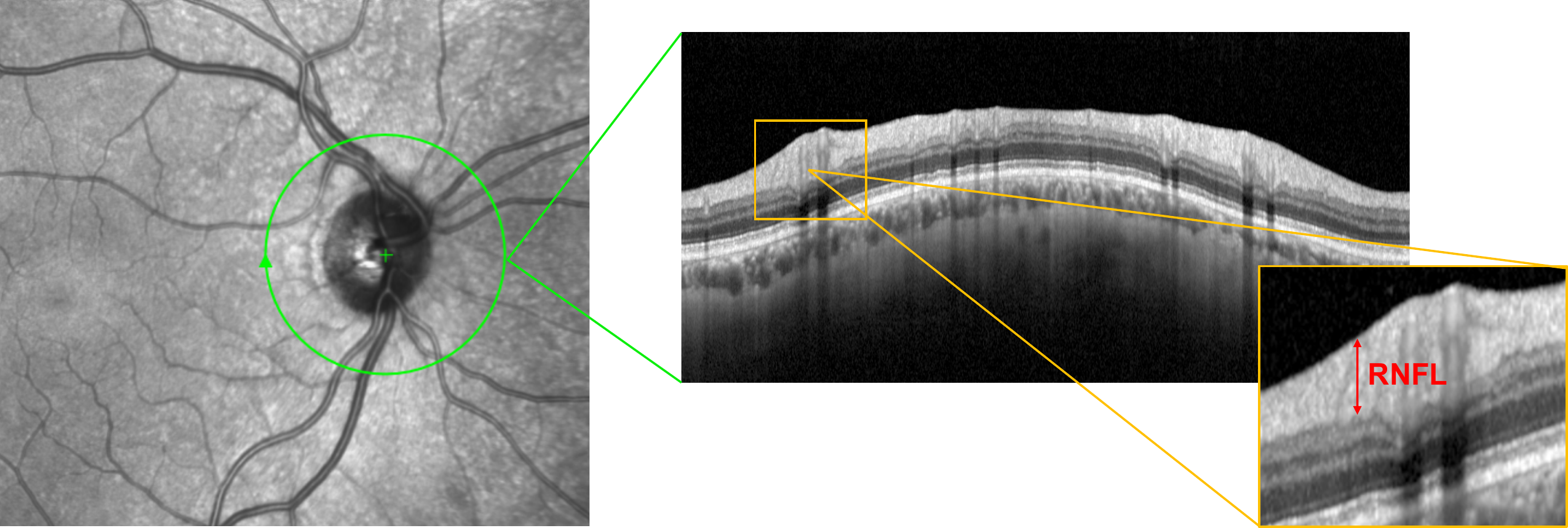}
\caption{B-scan around the retinal ONH corresponding to a circumpapillary OCT image. RNFL is highlighted in red.}
\label{fig:circum}
\end{figure}

Currently, a complete glaucoma study usually includes medical history, fundus photography, visual field (VF) analysis, tonometry and optic nerve imaging tests such as optical coherence tomography (OCT). Most of the state-of-the-art studies addressed the glaucoma detection via fundus image analysis, making use of visual field tests and relevant parameters like the intraocular pressure (IOP) \cite{kim2017, wang2019}. Specifically, J. Gómez-Valverde et al. \cite{gomez2019automatic} performed a comparison between convolutional neural networks (CNNs) trained from scratch and using fine-tuning techniques. Also, the authors in \cite{shibata2018development,christopher2018performance} considered the use of transfer learning and fine-tuning methods applied to very popular state-of-the-art network architectures to identify glaucoma on fundus images. Other studies such as \cite{muhammad2017hybrid, thakoor2019} carried out a combination between OCT B-scans and fundus images to obtain an RNFL thickness probability map which was used as an input to the CNNs. In this paper, contrary to the studies of the literature, we propose an end-to-end system for glaucoma detection based only on raw circumpapillary OCT images, without using another kind of images or external expensive tests related to the VF and IOP parameters. 
It is important to highlight that circumpapillary OCT images as shown in Fig. \ref{fig:circum} correspond to circular scans located around the ONH, where rich information about different retinal layers structures can be found. Additionally, several studies claimed that circumpapillary retinal nerve fiber layer (RNFL) is essential to detect early glaucomatous damage \cite{reference_objetive1, reference_objetive2, reference_objetive3}. For that reason, one of the main novelties of this paper is focused on demonstrating that a single circumpapillary OCT image may be of great interest when carrying out an accurate glaucoma detection.

We propose two different data-driven learning strategies to develop computer-aided diagnosis systems capable of discerning between glaucomatous and healthy eyes just from B-scans around the ONH. Several CNNs trained from scratch and different fine-tuned state-of-the-art architectures were considered. Furthermore, we propose, for the first time in this kind of images, the class activation maps computation in order to compare the location information reported by the clinicians with the heat maps generated by the developed models. Heat maps allow highlighting the regions in which the networks pay attention to determine the class of each specific sample.




\section{Material} \label{sec: Material}

The experiments detailed in this paper were performed on a private database composed of 249 OCT images of dimensions $M\times N=496\times768$ pixels. In particular, 156 normal and 93 glaucomatous circumpapillary samples were analysed from 89 and 59 patients, respectively. Each B-scan was diagnosed by experts ophthalmologists from Oftalvist Ophthalmic Clinic. Note that \textit{Heidelberg Spectrallis} OCT system was employed to acquire the circumpapillary OCT images with an axial resolution of 4-5 $\mu$m.

\section{Methodology} \label{sec: Methodology}

\subsection{Data Partitioning} \label{subsec: Data_partitioning}
A data partitioning stage was carried out to divide the database into different training and test sets. Specifically, $\frac{4}{5}$ of the circumpapillary images, which corresponds to 73 glaucomatous and 124 normal samples, from 12 and 18 patients respectively, composed the training set, whereas the test set was defined by $\frac{1}{5}$ of the data (20 with glaucoma and 32 normal B-scans from 12 and 18 patients). In addition, for the training set, we also performed an internal cross-validation (ICV) stage to control the overfitting, as well as to select the best neural network hyper-parameters. Finally, the independent test set was used to evaluate the definitive predictive models, which were created using the entire training set.

\subsection{Learning from scratch} \label{subsec: From_scratch}

Similarly to the methodology exposed in \cite{gomez2019automatic}, we propose in this paper the use of shallow CNNs from scratch to address the glaucoma detection, taking into account the significant differences between our grey-scale circumpapillary OCT images and other large databases containing natural images, which are widely used for transfer-learning techniques.

During the internal cross-validation (ICV) stage, an empirical exploration was carried out to determine the best hyper-parameter combination in terms of minimisation of the binary cross-entropy loss function. Different network architectures composed of diverse learning blocks were developed. In particular, convolutional, pooling, batch normalisation and dropout layers were considered to address the feature extraction stage. The variable components of each layer, such as the convolutional filters, pooling size, dropout coefficients, as well as the number of convolutional layers in each block were optimised during the experimental phase.  
Regarding the top model, the use of flatten, dropout and fully-connected layers with a different number of neurons was studied. Also, global max and global average pooling layers were analysed in order to reduce the number of trainable parameters.
Moreover, we implemented an optimal weighting factor of [1.35, 0.79] during the training of the models to alleviate the unbalanced problem between classes. 

After the ICV stage, the best CNN architecture was found using four convolutional blocks, as it is detailed in Table \ref{CNN_from_scratch}. It is remarkable the use of the global max-pooling (GMP) layer applied in the last block, which allows extracting the maximum activation of each convolutional filter before the classification layer. Also, note that batch normalization and dropout layers were not used because no improvement was reported during the validation phase. Only a dense layer with a \textit{softmax} activation and 2 neurons, corresponding to glaucoma and healthy classes, was defined. 

\begin{table}[htbp]
\caption{Proposed CNN architecture trained from scratch.}
\label{CNN_from_scratch}
\renewcommand{\arraystretch}{1.1} 
\setlength\tabcolsep{8 pt} 
\small
\begin{center}
\begin{tabular}{ccc}
\hline
 \textbf{Layer name} & \textbf{Output shape}                & \textbf{Filter size}            \\ \hline
Input layer         & 496 x 768 x 1                        & N/A                             \\
Conv1\_1            & 496 x 768 x 32                       & 3 x 3 x 32                      \\
MaxPooling          & 248 x 384 x 32                       & 2 x 2 x 32                      \\ 
Conv2\_1            & 248 x 384 x 64                       & 3 x 3 x 64                      \\ 
MaxPooling          & 124 x 192 x 64                       & 2 x 2 x 64                      \\ 
Conv3\_1            & 124 x 192 x 128                       & 3 x 3 x 128                      \\
MaxPooling          & 62  x 96  x 128                       & 2 x 2 x 128                      \\ 
Conv4\_1            & 62  x 96  x 256                      & 3 x 3 x 256                     \\
MaxGlobalPool       & 256                                  & N/A                             \\ 
Dense (softmax)     & 2                                    & N/A                             \\ \hline
\end{tabular}
\end{center}
\end{table}

The optimal hyper-parameters combination was achieved by training the CNNs during 150 epochs, using Adadelta optimizer with a learning rate of 0.05 and a batch size of 16. It should be noticed that we also proposed the use of data augmentation (DA) techniques \cite{wong2016} to elucidate how important is the creation of artificial samples when addressing small databases. Specifically, a factor ratio of 0.2 was applied here to perform random geometric and dense elastic transformations from the original images.

\subsection{Learning by fine tuning} \label{subsec: Deep_learning}

Deeper architectures networks could improve the models' performance, but a large number of images annotated by experts would be necessary for training a deep CNN from scratch. For this reason, we propose in this section the use of fine-tuning techniques \cite{hoo2016}, which allows training CNNs with greater depth using the weights pre-trained on large databases, without the need to train from scratch. In particular, we applied a deep fine-tuning \cite{tajbakhsh2016} strategy to transfer the wide knowledge acquired by several state-of-the-art networks, such as VGG16, VGG19, InceptionV3, Xception and ResNet, when they were trained on the large \textit{ImageNet} data set. Attending to the small database used in this work, only the coefficients of the last convolutional blocks (4 and 5) were retrained with the specific knowledge corresponding to the circumpapillary OCT images. The rest of coefficients were frozen with the values of the weights pre-trained with 14 million of natural images contained in \textit{Imagenet} database.

\begin{figure}[h]
\centering
\includegraphics[width=8.5cm]{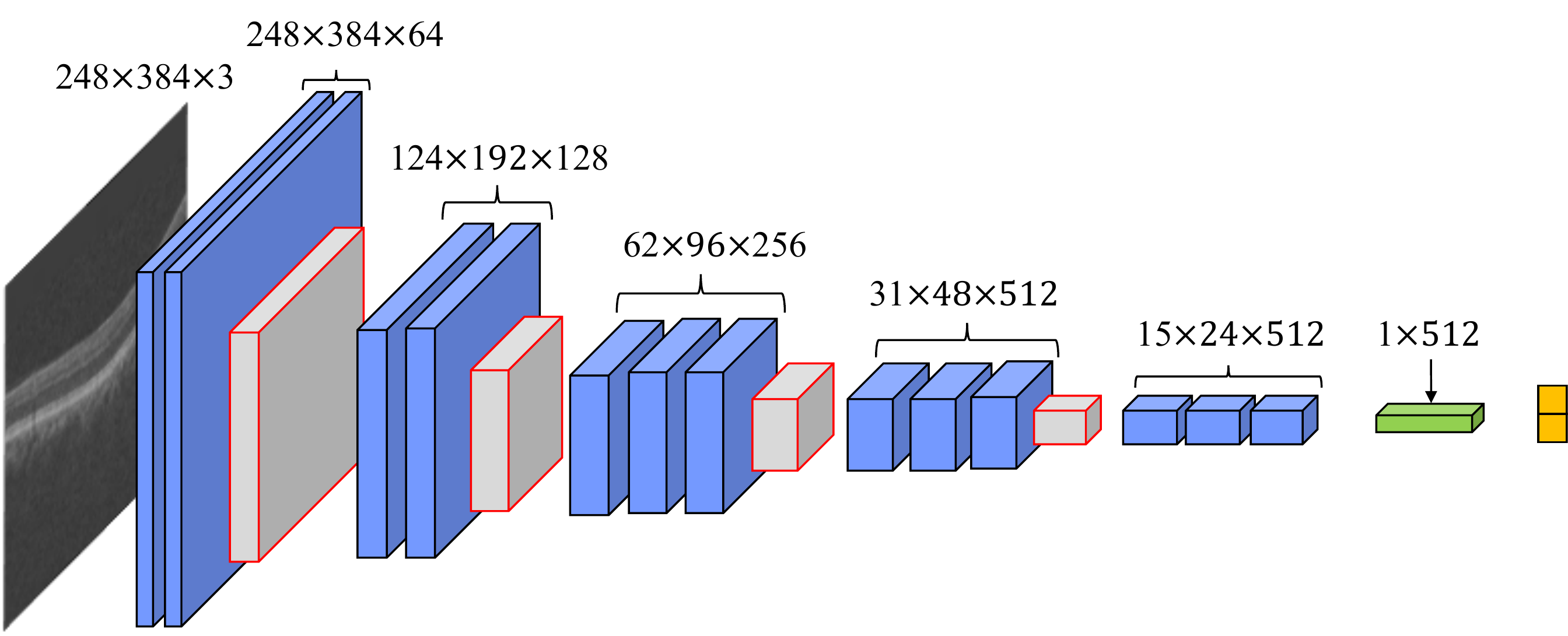}
\caption{Network architecture used to discern between glaucomatous and healthy OCT samples by fine-tuning the VGG16 base model. Note that numeric values of the filters are correctly defined in the image, although they do not correspond to the representation size of the boxes due to space problems.}
\label{fig:VGG16}
\end{figure}

Additionally, similarly to the proposed learning from scratch strategy, an empirical exploration of different hyper-parameters and top-model architectures was considered for all networks. It is important to notice that InceptionV3, Xception and ResNet architectures reported a poor performance due to their extensive depth (42, 36 and 53 convolutional layers, respectively). However, the family of VGG architectures achieved the best performance, in line with the findings in the literature \cite{gomez2019automatic}. Specifically, VGG16 base model is composed of five convolutional blocks according to Fig. \ref{fig:VGG16}, where blue boxes correspond to convolutional layers activated with \textit{ReLu} functions and red-grey boxes represent max-pooling layers. VGG19 base model is composed of the same architecture, but including an extra convolutional layer in the last three blocks. 

A top model composed of global max pooling and dropout layers with a coefficient of 0.4, followed by a softmax layer with two neurons, provided the best model performance when VGG architectures were fine-tuned (see Fig. \ref{fig:VGG16}). Regarding the selection of hyper-parameters combination, Adadelta optimizer with a learning rate of 0.001 reported the best learning curves when the model was forward, and backward, propagated during 125 epochs with a batch size of 16, trying to minimise the binary cross-entropy loss function.  

Note that an initial down-sampling $\times0.5$ of the original images was necessary to alleviate the GPU memory problems during the training phase. Besides, replicating $\times3$ the channels of the grey-scale was necessary to adapt the input images in order to fine tune the CNNs. Data augmentation (DA) techniques with a factor of 0.2 were also considered.

\section{Results and discussion} \label{sec: Results}

\subsection{Validation results}

In this stage, we present the results achieved during the ICV stage for each of the proposed CNNs. We expose in Table \ref{valRes_fromScratch} a comparison of the CNNs trained from scratch, in terms of mean $\pm$ standard deviation. Several figures of merit are calculated to evidence the differences between using or not data augmentation (DA) techniques. In particular, sensitivity (SN), specificity (SPC), positive predictive value (PPV), negative predictive value (NPV), F-score (FS), accuracy (ACC) and area under the ROC curve (AUC) are employed. 

\begin{table}[h]
\caption{Classification results reached during the ICV stage from the proposed CNNs trained from scratch.}
\label{valRes_fromScratch}
\setlength\tabcolsep{8 pt}
\small
\begin{center}
\begin{tabular}{cccc}
\hline
\multicolumn{1}{l}{}{} & \textbf{Without DA}       & \textbf{With DA} \\
\hline
\textbf{SN}         & $0.7657 \pm 0.2032$         &                  $\textbf{0.8771}\pm\textbf{0.1281}$\\
\textbf{SPC}        & $\textbf{0.9270} \pm \textbf{0.1302}$  &      $0.8047 \pm 0.1514$ \\
\textbf{PPV}        & $\textbf{0.8721} \pm \textbf{0.0662}$  &      $0.7477\pm 0.14061$  \\
\textbf{NPV}        & $0.8808 \pm  0.0971$                   &      $\textbf{0.9224} \pm \textbf{0.0678}$ \\
\textbf{FS}         & $\textbf{0.8016} \pm \textbf{0.1309}$  &      $0.7980\pm 0.10745$   \\
\textbf{ACC}        & $\textbf{0.8679} \pm \textbf{0.0781}$  &      $0.8315 \pm 0.0985$ \\
\textbf{AUC}        & $0.9152 \pm 0.0490$     &                     $\textbf{0.9319} \pm \textbf{0.0386}$\\
\hline
\end{tabular}
\end{center}
\end{table}

Significant differences between CNNs trained with and without data augmentation techniques can be appreciated in Table \ref{valRes_fromScratch}, especially related to the sensitivity and specificity metrics. 
Worth noting that the learning curves relative to the CNN trained without implementing DA algorithms reported slight overfitting during the validation phase. This fact is evidenced in the high sensitivity standard deviation of the model.

Additionally, we also detail in Table \ref{valtResults_fineTuning} the validation results achieved from the fine-tuned VGG networks, since they provided a considerable outperforming with respect to the rest of state-of-the-art architectures during the ICV stage. Specifically, VGG16 reaches better results for all figures of merit, although both architectures report similar behaviour. In comparison to the CNNs trained from scratch, VGG16 provides the best model performance too. 

\begin{table}[h]
\caption{Results comparison between the best fine-tuned CNNs proposed during the validation phase.}
\label{valtResults_fineTuning}
\setlength\tabcolsep{8 pt}
\small
\begin{center}
\begin{tabular}{cccc}
\hline
\multicolumn{1}{l}{}{} & \textbf{VGG16}       & \textbf{VGG19} \\
\hline
\textbf{SN}         & $\textbf{0.7800}   \pm \textbf{0.1302}$       & $0.7400 \pm 0.1462$ \\
\textbf{SPC}        & $\textbf{0.9677} \pm \textbf{0.0334}$         & $0.9597\pm 0.0283$\\
\textbf{PPV}        & $\textbf{0.9401} \pm  \textbf{0.0643}$      & $0.9180 \pm 0.0602$ \\
\textbf{NPV}        & $\textbf{0.8864} \pm \textbf{0.0662}$        & $0.8670\pm 0.0692$  \\
\textbf{FS}         & $\textbf{0.8466} \pm  \textbf{0.0720}$       & $0.8131\pm 0.0936$   \\
\textbf{ACC}        & $\textbf{0.8984} \pm  \textbf{0.0468}$      & $0.8786 \pm 0.0563$ \\
\textbf{AUC}        & $\textbf{0.9463} \pm  \textbf{0.0339}$     & $0.9416 \pm 0.0501$\\
\hline
\end{tabular}
\end{center}
\end{table}


\subsection{Test results}
 
\begin{figure}[h]
\begin{center}
\includegraphics[height=3cm, width=7.25cm]{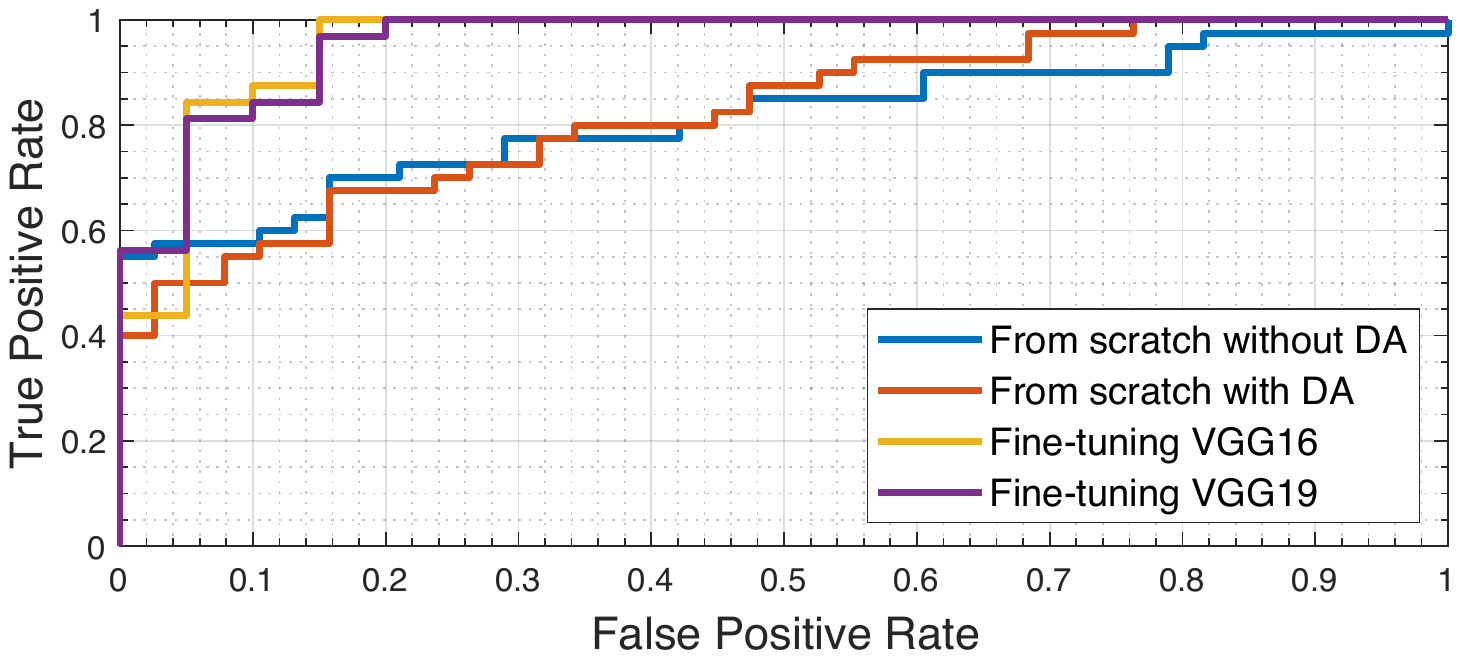} \\
\end{center}
\caption{ROC curves corresponding to the prediction results reached from the different proposed CNNs.}
\label{testROCs}
\end{figure}

In order to provide reliable results, an independent test set was used to carry out the prediction stage. Table \ref{testResults} shows a comparison between all proposed deep-learning models to evaluate their prediction ability by means of different figures of merit. Additionally, we expose in Fig. \ref{testROCs} the ROC curve relative to each proposed CNN to visualise the differences.

\begin{table}[h]
\caption{Classification results achieved during the prediction stage from the proposed CNNs trained from scratch (FS) and fine-tuning the VGGs network architectures.}
\label{testResults}
\small
\begin{center}
\begin{tabular}{ccccc}
\hline
\multicolumn{1}{l}{}{} & \textbf{FS without DA}       & \textbf{FS with DA}  & \textbf{VGG16}       & \textbf{VGG19}\\
\hline
\textbf{SN}          & 0.7632           & 0.7895    & \textbf{0.8510}       & \textbf{0.8510}  \\
\textbf{SPC}         & 0.7250           & 0.6750    & 0.9064                & \textbf{0.9688} \\
\textbf{PPV}         & 0.7250           & 0.6977    & 0.8490                & \textbf{0.9444}   \\
\textbf{NPV}         & 0.7632           & 0.7714    & 0.9063                & \textbf{0.9118}   \\
\textbf{FS}          & 0.7436           & 0.7407    & 0.8500                & \textbf{0.8947}    \\
\textbf{ACC}         & 0.7436           & 0.7308     & 0.8846               & \textbf{0.9230}   \\
\textbf{AUC}         & 0.8132           & 0.8230     & 0.9578               & \textbf{0.9594} \\
\hline
\end{tabular}
\end{center}
\end{table}

Test results exposed in Fig. \ref{testResults} are in line with those achieved during the validation phase. However, due to the randomness effect of the data partitioning (which is accentuated in small databases), significant differences may exist in the prediction of each subset. This fact mainly affects to the CNNs trained from scratch because all the weights of the network were trained with the images of a specific subset, whereas the proposed fine-tuned architectures keep most of the weights frozen. Regarding the ROC curves comparison, Fig. \ref{testROCs} shows that fine-tuned CNNs report a significant improvement in relation to the networks trained from scratch.

It is important to remark that an objective comparison with other state-of-the-art studies is difficult because there are no public databases of circumpapillary OCT images. Additionally, each group of researchers addresses glaucoma detection using a different kind of images. Notwithstanding, we detail a subjective comparison with other works based on similar methodologies applied to fundus images. In particular, \cite{gomez2019automatic} fine-tuned the VGG19 architecture and achieved an AUC of 0.94 predicting glaucoma. Also, \cite{christopher2018performance} reached an AUC of 0.91 applying transfer learning techniques to the ResNet architecture. Otherwise, authors in \cite{chen2015glaucoma} proposed a CNN from scratch obtaining AUC values of 0.83 and 0.89 from two independent databases. Basing on this, the proposed learning methodology exceeds the state-of-the-art results, achieving an AUC of 0.96 during the prediction of the test set.

\BlankLine
\textbf{Class Activation Maps (CAMs)}

\begin{figure}[h]
\begin{center}
\begin{tabular}{cc}
\includegraphics[height=4.2cm,width=4cm]{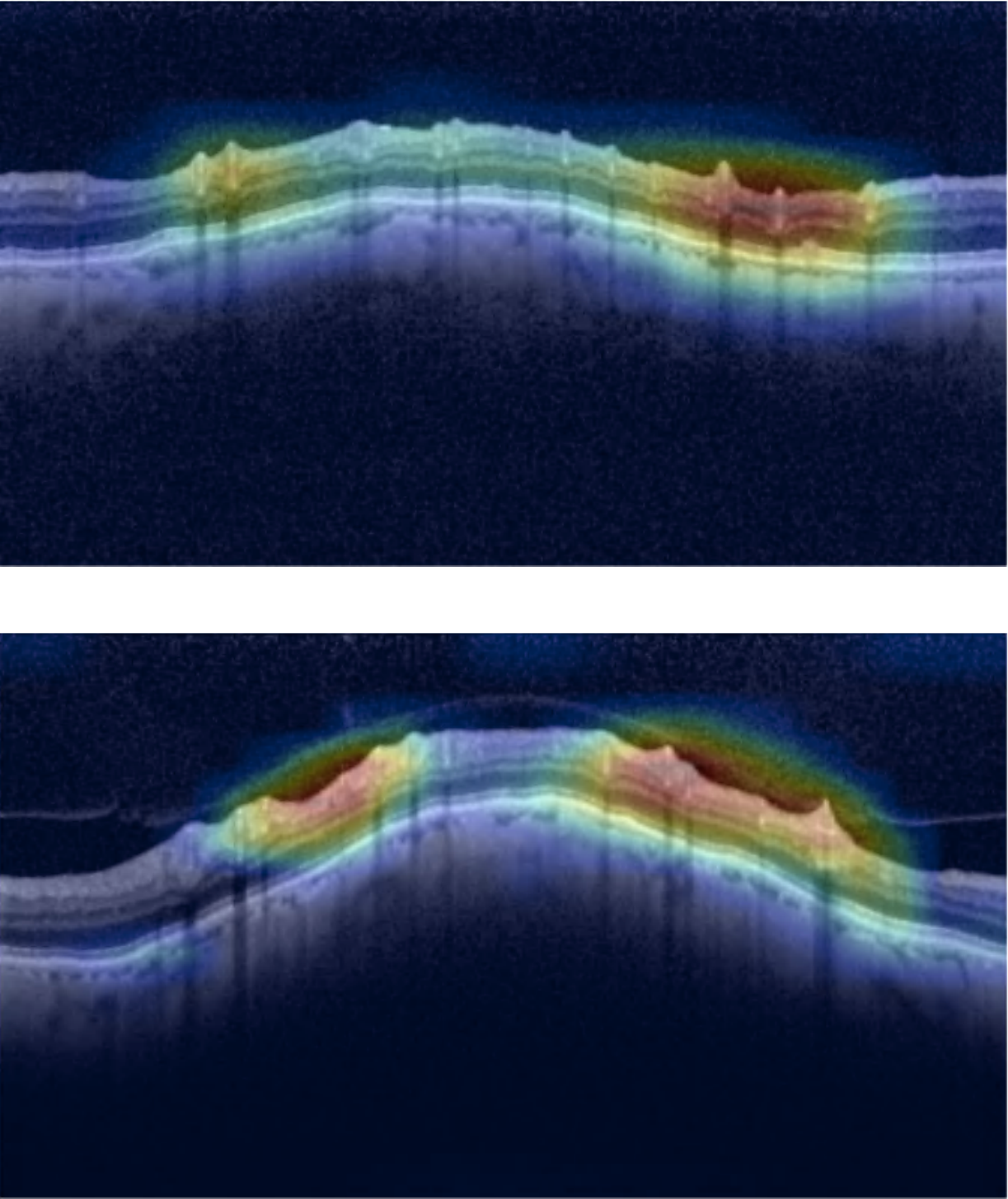} &
\includegraphics[height=4.2cm, width=4cm]{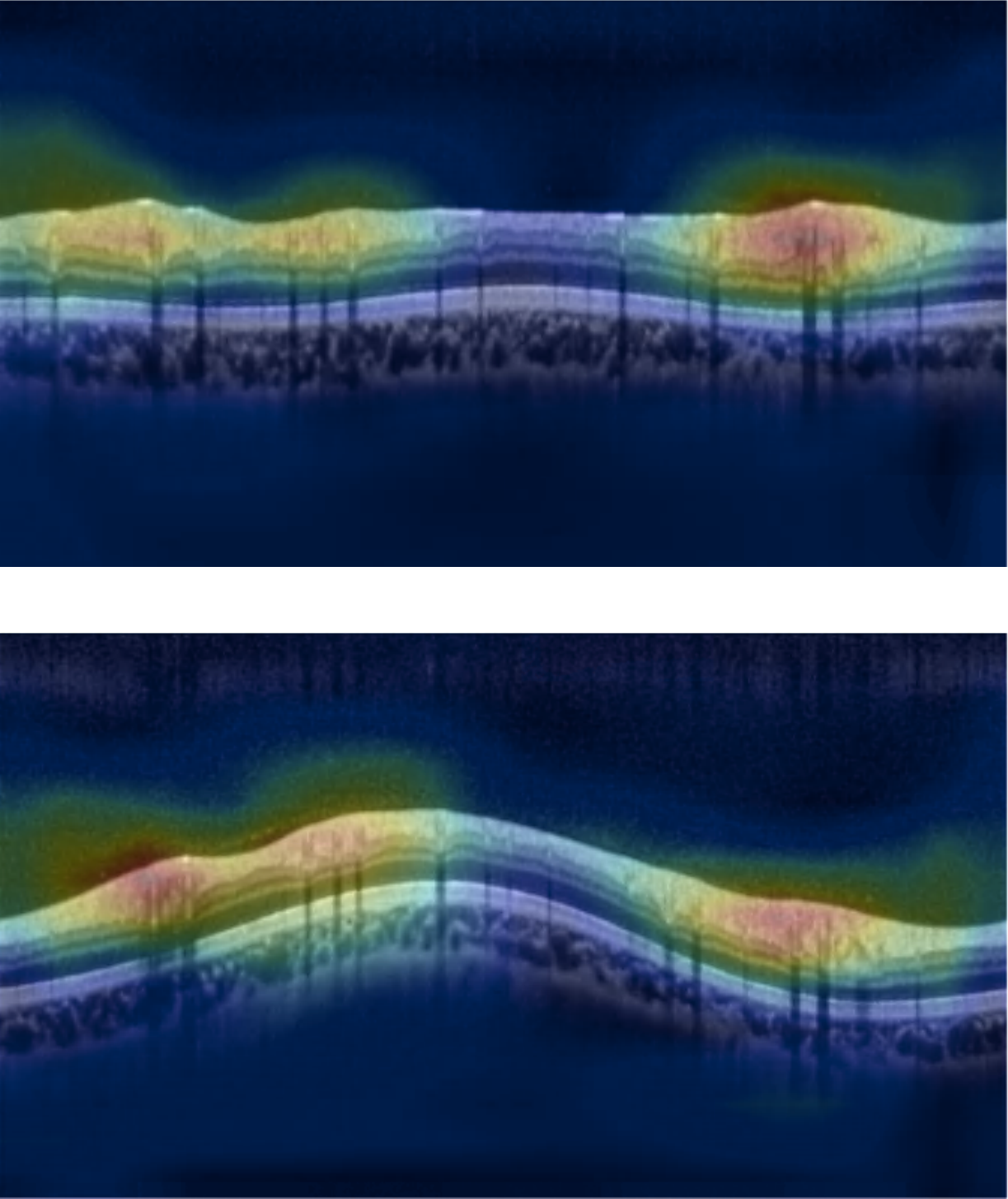} \\
(a) &
(b) \\
\end{tabular}
\end{center}
\caption{Heat maps extracted from the CAMs computation for (a) glaucomatous and (b) healthy circumpapillary images.}
\label{Class_Activation}
\end{figure}

We compute the class activation maps to generate heat maps highlighting the interesting regions in which the proposed model is paying attention to determine the class of each specific circumpapillary OCT image. In Fig. \ref{Class_Activation}, we expose the CAMs relative to random specific glaucomatous and normal samples in order to elucidate what is VGG19 taking into account to discern between classes.

The findings from the CAMs are directly in line with the reported by expert clinicians, who claim that a thickening of the RNFL is intimately linked with healthy patients, whereas a thinning of the RNFL evidence a glaucomatous case. That is just what heat maps in Fig. \ref{Class_Activation} reveal. Therefore, the results suggest that the proposed circumpapillary OCT-based methodology can provide a great added value for glaucoma diagnosis taking into account that information similar to that of specialists is reported by the model without including any previous clinician knowledge.

\section{Conclusion} \label{sec: Conclusion}

In this paper, two different deep-learning methodologies have been performed to elucidate the added value enclosed in the circumpapillary OCT images for glaucoma detection. The reported results suggest the fine-tuned VGG family of architectures as the most promising networks. The extracted CAMs evidence the great potential of the proposed model since it is able to highlight areas such as the RNFL, in line with the clinical interpretation. In future research lines, external validation of the proposed strategy with large databases is considered.

\newpage


\balance
\bibliographystyle{IEEEbib}
\bibliography{strings,refs}

\end{document}